\documentclass[submission,copyright,creativecommons]{eptcs}
\providecommand{\event}{HSB 2012} 
\usepackage{color}
\usepackage{array}
\usepackage{amsmath,amssymb}
\usepackage{graphicx}
\usepackage{faktor}
\usepackage[caption=false,font=footnotesize]{subfig}

\newtheorem{thm}{Theorem}

\newtheorem{prop}[thm]{Proposition}
\newtheorem{defn}[thm]{\ \\Definition}

\title{Disease processes as hybrid dynamical systems}
\author{Pietro Li\`{o}
\institute{Computer Laboratory\\University of Cambridge\\Cambridge, UK}
\email{pl219@cam.ac.uk}
\and
Emanuela Merelli \qquad\qquad Nicola Paoletti
\institute{School of Science and Technology\\Computer Science Division, University of Camerino\\Camerino, IT}
\email{\quad emanuela.merelli@unicam.it \quad\qquad nicola.paoletti@unicam.it}
}
\def\titlerunning{Disease processes as hybrid dynamical systems}
\def\authorrunning{P. Li\`{o}, E. Merelli \& N. Paoletti}
\begin{document}
\maketitle

\begin{abstract}
We investigate the use of hybrid techniques in complex processes of infectious diseases. Since predictive disease models in biomedicine require a multiscale approach for understanding the molecule-cell-tissue-organ-body interactions, heterogeneous methodologies are often employed for describing the different biological scales. Hybrid models provide effective means for complex disease modelling where the action and dosage of a drug or a therapy could be meaningfully investigated: the infection dynamics can be classically described in a continuous fashion, while the scheduling of multiple treatment discretely. We define an algebraic language for specifying general disease processes and multiple treatments, from which a semantics in terms of hybrid dynamical system can be derived. Then, the application of control-theoretic tools is proposed in order to compute the optimal scheduling of multiple therapies. The potentialities of our approach are shown in the case study of the SIR epidemic model and we discuss its applicability on osteomyelitis, a bacterial infection affecting the bone remodelling system in a specific and multiscale manner. We report that formal languages are helpful in giving a general homogeneous formulation for the different scales involved in a multiscale disease process; and that the combination of hybrid modelling and control theory provides solid grounds for computational medicine.
\end{abstract}

\section{How many scales does it take to describe an epidemics?}
\begin{figure}
\centering
\includegraphics[width=\textwidth]{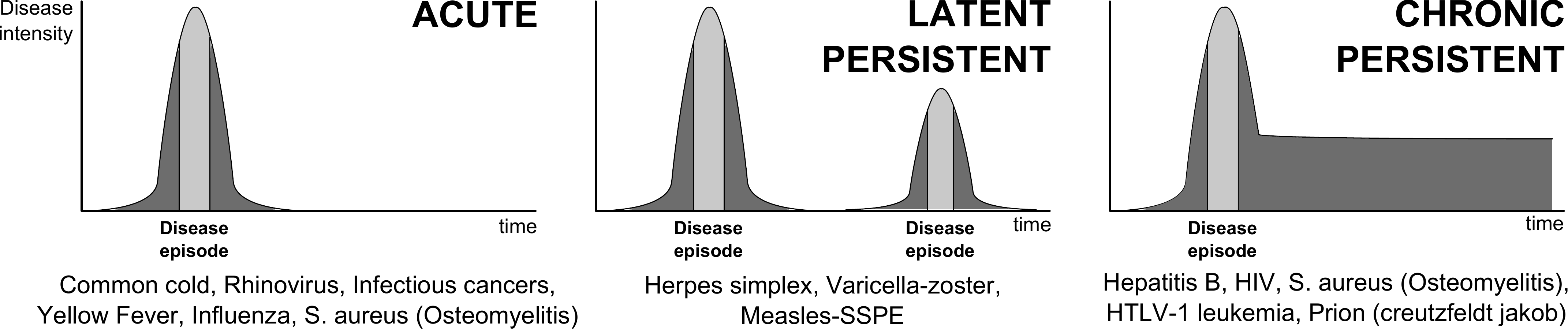}
\caption{Simplified classification of infectious diseases. The disease intensity follows a Gaussian curve approximately partitioned in three stages. In the first one, the infection starts but any symptom is reported. In the second one called ``disease episode'' symptoms are reported and the disease can be transmitted to other individuals. In the last stage, the disease is over if it is acute or latent persistent. However some diseases can become chronic.}
\label{fig:classification}
\end{figure}

In this work we investigate multi-methodology approaches for the modelling of complex infectious diseases and for finding the best strategy of intervention, able to avoid or limit the spread of the disease and its effect on the affected organisms. Infectious diseases can be broadly classified into three groups (Fig.~\ref{fig:classification}): acute, latent persistent and chronic persistent. Acute diseases like the common cold, the Rhinovirus, the Yellow Fever, the Influenza or some strains of the Staphylococcus Aureus (Osteomyelitis) are characterized by a single disease episode after which they do not occur anymore. Latent persistent ones like the Herpes simplex, the Varicella-zoster or the Measles-SSPE arises also after the first disease episode and a non-infectious and latent period. Chronic persistent diseases (e.g.\ Hepatitis B, HIV, HTLV-1 leukemia, chronic Osteomyelitis) can protract their effects on the host organism for several years. 

Epidemic modelling is one of the most established tools for predicting the progress and the spread of a disease in large populations, which are commonly divided into compartments:
\textbf{S}usceptible, \textbf{E}xposed, \textbf{I}nfected and \textbf{R}ecovered. According to the possible flows among such compartments, different variants arises, like the SIR (Susceptible$\rightarrow$Infected$\rightarrow$Recovered) model~\cite{kermack1932contributions}, SIS, SIRS, SEIR and SEIRS. Further compartments can be taken into account like the population of immune infants, usually indicated with M. Similar compartments can be identified not only at the human population level, but also at the cellular population level, where the pathogen typically acts by infecting susceptible cells. As a matter of fact, infections are characterized by multiscale dynamics that affect the organism at multiple levels in the biological hierarchy, as shown on the left side of Fig.~\ref{fig:multiple_disease}: at the \textit{intracellular} level, in case of infection of a cell by a pathogen; at the \textit{intercellular/cellular population} level, in case of infection of susceptible cells by infected cells; at the \textit{tissue} level, when multi-cellular ensembles are involved; at the \textit{organ/individual} level, when infection spreads to other parts of the body; and at the \textit{human population} level, if the disease is transmitted among individuals. A further scale above the human population one could be the consciousness level for which several psychological and cognitive behaviour models exists. Just consider that psychological states like fear or stress can even affect (negatively) the immune system: a demonstration that all the scales in the biological hierarchy are intimately connected to each other. Their inherent multiscale dynamics make the modelling of infective processes a challenging and intriguing field of research. Indeed, pathogens and infective agents generally act at the cellular scale, but they need to exploit mechanisms at the human population level to spread and survive in other host organisms.

\subsection{Formal languages, hybrid modelling and control in multiscale infectious diseases}
\begin{figure}
\centering
\includegraphics[width=0.7\textwidth]{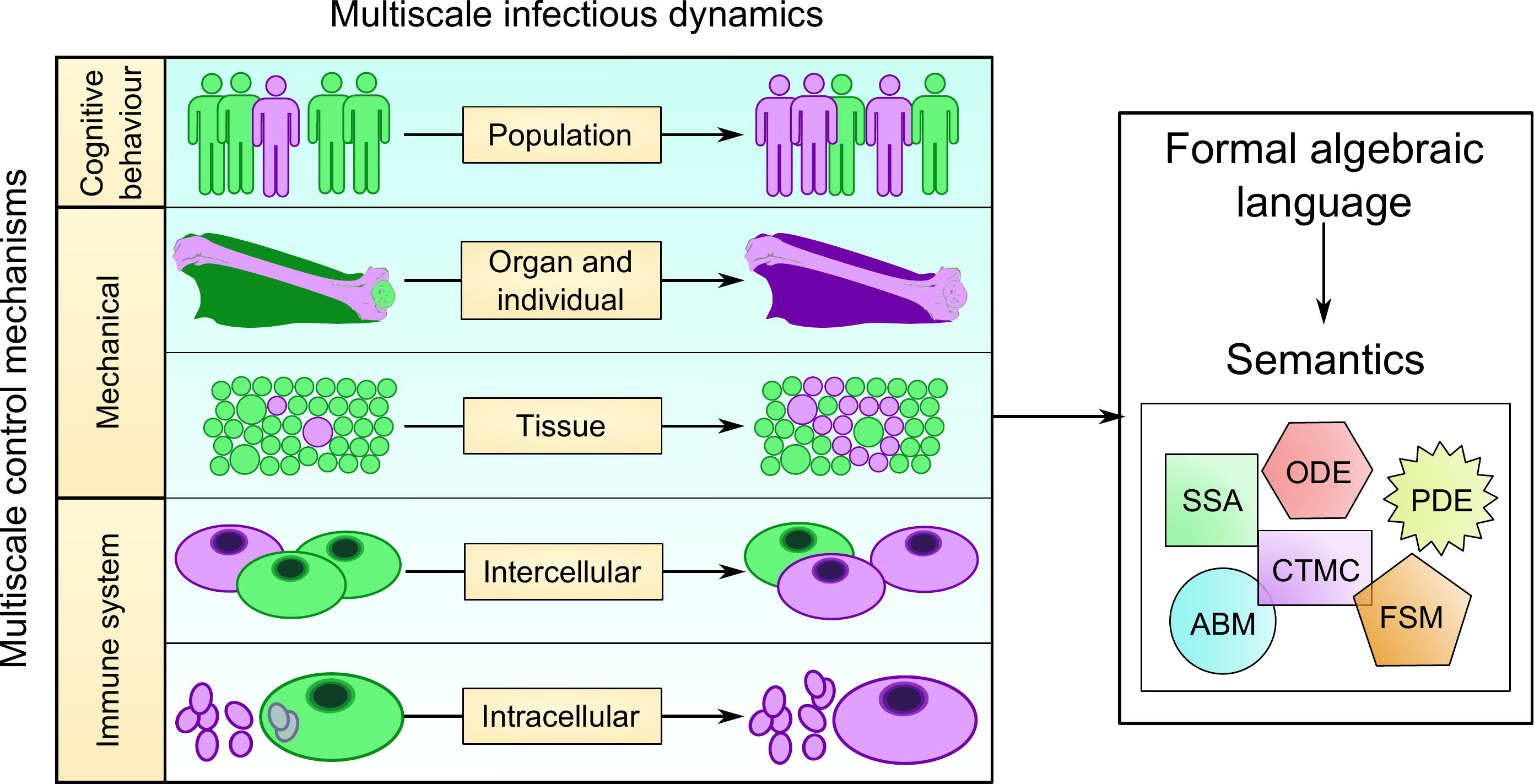}
\caption{Modelling of multiscale infective dynamics. Infections involve the intracellular level (a cell infected by a pathogen); the intercellular level (infection among cells); the tissue level (infection among collections of cells); the organ level (infection among different parts of the body); and the human population level (transmission among individuals). Each biological scale can be implemented with different modelling techniques like Gillespie's SSA, Continuous Time Markov Chains (CTMC), ordinary/partial differential equations (ODE/PDE), agent-based models (ABM), or finite state machines (FSM). Such heterogeneous semantics can be wrapped by a common formal language.}
\label{fig:multiple_disease}
\end{figure}
We believe that hybrid modelling could help in unravelling the complexity of the multiscale dynamics occurring in such diseases. We refer to 'hybrid modelling' not just with its classical meaning, i.e.\ the modelling of those systems characterized by the co-existence of continuous and discrete dynamics, but also with a methodological meaning. In multiscale systems different scales are typically approached with different methodologies, thus leading to the co-existence of heterogeneous modelling techniques, in other words to a methodologically hybrid modelling approach. The cellular level is typically represented with agent-based or ordinary differential equation models, while the tissue and organ levels are often described using image-based finite element modelling (partial differential equations). In the context of disease modelling the determination of suitable intervention strategies, including drug and therapy administration, adds another level of description that affects multiple biological scales. 

This kind of integrative models are composed of single-scale models, describing the biological process at different characteristic space-time scales, and scale bridging models, which define how the single-scale models are coupled to each other~\cite{viceconti2011multiscale}. Higher level languages and formalisms can help in giving a general homogeneous formulation for the different scales in a multiscale biological system. In general, that formal description cannot be directly executed or simulated, but it is able to support multiple semantics (e.g. transition systems, differential equations and Markov chains). In particular process-algebraic languages are a formal notation initially developed for modelling software systems, but in the last decade it has been extensively used and extended in order to describe biological systems. In a seminal paper~\cite{cardelli2008processes}, Cardelli showed that a subset of stochastic CCS is powerful enough to encode both systems of reactions in the stoichiometric form and systems of ordinary differential equations. Additionally, a recent work of the authors~\cite{paoletti2011,	paoletti2011spec} demonstrates the effectiveness of a hybrid approach where a process-algebraic specification level is translated into a runnable stochastic agent-based model in the bone remodelling case study, while in~\cite{bartocci2012tcsb} the same case study is approached with different semantics according to the biological property to analyse.

Consequently, formal languages could represent a wrapping language able to homogeneously describe the different scales of a complex multiscale system, where each level can be instantiated into a runnable model according to the most suitable semantics. Figure~\ref{fig:multiple_disease} sketches the idea of the heterogeneous semantics associated to a multiscale model of infection through a common formal language. Furthermore, the importance of being hybrid in the classical sense is demonstrated by several biological evidences. For instance, genetic regulatory networks naturally exhibits hybrid dynamics, for which the continuous concentration of proteins is interrupted by the discrete switches dictated by changes in gene expression~\cite{batt2005qualitative,grosu2011cardiac}. Moreover in population biology it happens that some species are present in high concentration, so they can be approximately modelled by continuous variables; conversely, small populations are opportunely modelled as discrete stochastic variables~\cite{lapin2011shave}.

Another crucial aspect is related to the self-regulation and control of biological systems: in normal conditions, biological entities and functions are self-regulated and several multiscale control mechanisms naturally exists. One of the most striking example is the control operated by the immune system that protect the organism in case of disease. In fact, lymphocytes are able to detect the presence of a pathogen and produce an appropriate immune response by secreting immunoglobulins. However, in many cases of severe diseases the immune system cannot apply an effective control anymore. Drugs and therapies represent a form of control which is external to the organism and is extremely significant when dealing with models of diseases that can be limited by appropriate medical interventions. The control operated by medical therapies can be much more effective but needs to formulate control laws that take into account their (negative) impact on the organism due to possible side effects. While the drug administration strategy is determined at the human population level, the therapy affects multiple lower scales (in most of the cases, the cellular and subcellular scale). Other examples of multiscale control are the mechanical control at the organ and tissue level that regulates the functioning of bone cells in the bone remodelling process; or in an infection scenario, the cognitive behaviour at the human level that avoids or limits the interactions with other individuals possibly affected by the disease. A work by Bagnoli et al.~\cite{bagnoli2007risk} embeds this aspect of human behaviour into a SIR epidemic model by allowing individuals to perceive the risk of being infected by ill neighbours. Therefore, the use of control-theoretic tools seems promising for describing the biological mechanisms of multiscale self-regulation and self-adaptiveness, as well as for implementing externally imposed control mechanisms. 

In this work we present a general model for complex multiscale infectious diseases and for scheduling optimal medical treatments. We employ a hybrid modelling approach both in the classical and in the methodological sense. Indeed, we formulate the disease process by means of a high-level process algebra called D-CGF, from which a hybrid (in the classical sense) semantics can be derived. This semantics is given in terms of a hybrid dynamical system, where the cell population and the infection dynamics are described in a continuous fashion, while the dosage of multiple therapies is implemented as discrete switches. Then, we propose the application of model predictive control (MPC) tools for computing the optimal scheduling of multiple therapies. The potentialities of our approach are shown in the case study of the SIR epidemic model and we discuss its applicability on the case study of osteomyelitis, a bacterial infection affecting the bone remodelling system.

The paper is organized as follows. Section~\ref{sect:general_model} presents the algebraic language D-CGF for generic disease processes and its semantics in terms of hybrid dynamical systems, by means of the SIR example. Section~\ref{sect:sim} reports the optimal controlled solutions for scheduling multiple therapies in the SIR model and show how this approach can be useful to the osteomyelitis case study. Discussions and conclusions are given in Section~\ref{sect:conclusions}. 

\section{General process-algebraic formulation of infectious diseases}\label{sect:general_model}
In this part we propose a generic formulation of complex infectious diseases, by defining a variant of the Chemical Ground Form (CGF) stochastic process algebra. Stochastic process algebras extend classical ones with quantitative information in form of action rates and have associated a Continuous Time Markov Chains (CTMC) semantics from which a continuous and less computationally demanding approximation in terms of ODE can be derived. The multiplicity of semantics it supports has made formal process-algebraic languages one of the most used tools in biological modelling. A work by Cardelli~\cite{cardelli2008processes} shows how several encodings can be implemented among specifications in the CGF algebra (a subset of stochastic CCS), systems of reactions in the stoichiometric form and ODE systems. Bortolussi and Policriti~\cite{bortolussi2009hybrid} extend this work by defining a semantics in terms of hybrid automata~\cite{henzinger1996theory}, defining the requisites according to which a set of processes can be treated as discrete control states, while others have the usual continuous interpretation. Several other methods for associating hybrid semantics to quantitative process algebras have been proposed, among which~\cite{galpin2009hype,bortolussi2010hybrid}.

The formal modelling language employed in this paper is called \textit{Disease Chemical Ground Form (D-CGF)}, and it is a variant of CGF for describing complex disease processes. Differently from~\cite{norman2003developing}, where a stochastic process algebra and its continuous semantics are used to describe a classical epidemic model, we attempt to provide a framework for more generic diseases and, although the syntax of D-CGF does not radically differ from that of CGF, we define a novel constructive procedure for deriving a semantics in terms of hybrid dynamical systems, by distinguishing the following two kinds of processes:
\begin{itemize}
\item \textit{Individuals}: standard CGF processes that belong to some species and that collectively represent the population in the disease model. Different species can describe for instance the various compartments of an epidemic model (susceptible, infected, and etc.), a pathogen, or a particular mutation. Individuals are interpreted as continuous variables in the hybrid semantics.
\item \textit{Therapies}: processes for modelling interventions on the disease scenario. They could represent for instance the dosage of a drug or a change in the environment that \textit{discretely} alter the disease dynamics. Based on the hybrid semantics of CGF~\cite{bortolussi2009hybrid}, therapy processes are subject to some restrictions in order to be interpreted as the discrete switches of the hybrid dynamical system.
\end{itemize}
D-CFG will be illustrated by means of an epidemic example, even if this language is suitable to describe also models of virus infection, infectious cancers or other kinds of diseases where discrete intervention policies has to be modelled.
\subsection{The Disease Chemical Ground Form}
Here we present the \textit{Disease Chemical Ground Form (D-CGF)}, a variant of the Chemical Ground Form (CGF)~\cite{cardelli2008computational}) targeted to describe complex models of disease in a process-algebraic fashion. The syntax of \textit{D-CGF} is given by the grammar in Table~\ref{tbl:dcgfSyntax}. A D-CGF model is given by a set of species $S$, an initial population $P$, a set of therapies $T$, and an initial combination of therapies $C$. The first notable difference from the algebra CGF is that in D-CGF models two disjoint sets of processes are distinguished:
\begin{itemize}
\item the set of species (a set of individual species definitions), the individual species and the population (the parallel composition of individuals); and
\item the set of therapy definitions, the therapy and the combination of therapies (the parallel composition of therapies).
\end{itemize}
\begin{table}
\centering
\begin{small}
\begin{tabular}{rll}
$S::=$&$\mathbf{0} \ | \ X = I, S$ & Set of species definition\\
$I::=$&$\mathbf{0} \ | \ \pi . P + I$ & Individual species\\
$P::=$&$\mathbf{0} \ | ( X \| P)$ & Population\\
$\pi::=$&$\tau^r \ | \ ? x^r \ | \ ! x^r$ & Actions\\
&&\\
$T::=$&$\mathbf{0} \ | \ U = R, T$ & Set of therapy definition\\
$R::=$&$\mathbf{0} \ | \ \pi . C + R$ & Therapy\\
$C::=$&$\mathbf{0} \ | ( U \| C)$ & Combination of therapies\\
&&\\
$D-CGF::=$&$(S, P, T, C)$ & A \textit{D-CGF} model
\end{tabular}
\end{small}
\caption{Syntax of D-CGF.}\label{tbl:dcgfSyntax}
\end{table}
Individual species and therapies are defined by the alternative composition ($+$) of action-prefixed terms. Actions $\pi \in \Pi$ are indexed by a quantity $r$ corresponding to the rate of the exponential distribution that determines their duration, and are the usual ones: $\tau^r$ (internal action), $? x^r$ (input action) and $! x^r$ (output action). The motivation for separating these two sets of processes is related to the hybrid semantics of D-CGF: species are interpreted as continuous variables, while therapies as discrete control states. In addition we will need to put some restrictions on the therapy processes, for formally justifying this separation. In the remainder of this section, a simple epidemic model will serve as the running example for illustrating our approach.
\subsubsection{D-CGF by example}
In this part, we illustrate some of the technical features of the algebra by means of an epidemic example: the SIR (Susceptible$\rightarrow$Infected$\rightarrow$Recovered) model~\cite{kermack1932contributions}. It describes the dynamics at the epidemiological level of a population consisting of individuals susceptible to the disease ($S$), those infected ($I$) and those recovered ($R$). Here we consider an open-population (i.e.\ with births and deaths) SIR model, given by the list of reactions in Table~\ref{tbl:sirmodel}.
\begin{table}
\centering
\begin{small}
\begin{tabular}{rl}
$S \rightarrow_b S + S$&\\
$I \rightarrow_b I + S$&Birth of susceptible\\
$R \rightarrow_b R + S$&\\
\noalign{\smallskip}
$S \rightarrow_\mu \emptyset$&Death of susceptible\\
$I \rightarrow_\mu \emptyset$&Death of infected\\
$R \rightarrow_\mu \emptyset$&Death of recovered\\
\noalign{\smallskip}
$S + I \rightarrow_\beta I + I$&Infection of susceptible\\
\noalign{\smallskip}
$I \rightarrow_\nu R$&Infected becomes recovered
\end{tabular}
\end{small}
\caption{Reactions in the SIR epidemic model with open population and no disease-dependent death rate.}
\label{tbl:sirmodel}
\end{table}

According to the conversion rules for CGF, we can translate this list of reactions in a D-CGF model $\mathcal{M}=(S, P, \mathbf{0}, \mathbf{0})$ with no therapies.
{\small
\[
S:\ 
\begin{array}{rl}
S = & \tau_{S_1}^b . (S \| S) \ + \ \tau_{S_2}^\mu . \mathbf{0}\ + \ ? i^\beta . I\\
I = & \tau_{I_1}^b . (I \| S) \ + \ \tau_{I_2}^\mu . \mathbf{0}\ + \ ! i^\beta . I\ + \ \tau_{I_3}^\nu . R\\
R = & \tau_{R_1}^b . (R \| S) \ + \ \tau_{R_2}^\mu . \mathbf{0}
\end{array}
\qquad
P: \ P_0,
\]
}\noindent where $P_0$ is the initial population. Given an action $\pi \in \Pi$, we denote with $react(\pi)$ the multiset of species consumed by $\pi$, with $prod(\pi)$ the multiset of those produced by $\pi$. $\#(X,P)$ denotes the number of $X$ occurring in the population $P$. $\Delta (\pi, X) = \#(X,prod(\pi)) - \#(X,react(\pi))$ denotes the net variation of $X$ due to $\pi$. For instance in the above example, $react(i) = \lbrace S, I\rbrace$, $prod(i) = \lbrace I, I\rbrace$, $\Delta (i, S) = -1$ and $\Delta (i, I) = 1$.

The procedure for extracting a system of ODEs requires to build the so-called \textit{stoichiometric matrix} $M$, which has one row for each species and one column for each action. A cell $M[X,\pi]$ takes the net variation of $X$ due to $\pi$: $\Delta (\pi, X)$. Then a \textit{rate vector} $\phi$ has to be defined in the following way. Let $r(\pi)$ denote the rate of action $\pi$. Then for each action $\pi \in \Pi$,
{\small
\[
  \phi[\pi] = \left\{ 
  \begin{array}{l l}
    0 & \quad \text{if } react(\pi)=\emptyset\\
    r(\pi)X & \quad \text{if } react(\pi)=\lbrace X\rbrace\\
    r(\pi)XY & \quad \text{if } react(\pi)=\lbrace X, Y\rbrace\\
    r(\pi)X(X-1) & \quad \text{if } react(\pi)=\lbrace X, X\rbrace\\
  \end{array} \right.
\]
}\noindent Finally, the differential equations are given by $\mathbf{\dot{X}} = M \cdot \phi$. The stoichiometric matrix, the rate vector and the resulting ODEs for the SIR example are shown below.
{\footnotesize
\[
M = \bordermatrix{~ &\tau_{S_1} & \tau_{I_1} & \tau_{R_1} & \tau_{S_2} & \tau_{I_2} & \tau_{R_2} & \tau_{I_3} & i \cr
S& 1 & 1 & 1 & -1 & 0 & 0 & 0 & -1 \cr
I& 0 & 0 & 0 & 0 & -1 & 0 & -1 & 1 \cr
R& 0 & 0 & 0 & 0 & 0 & -1 & 1 & 0 \cr}
\qquad
\phi = \bordermatrix{~ &\cr
\tau_{S_1}&  bS\cr
\tau_{I_1}&  bI\cr
\tau_{R_1}&  bR\cr
\tau_{S_2}&  \mu S\cr
\tau_{I_2}&  \mu I\cr
\tau_{R_2}&  \mu R\cr
\tau_{I_3}&  \nu I\cr
i&  \beta SI \cr
}
\qquad
\begin{matrix}
\dot{S} = bN - \beta S I - \mu S\\
\dot{I} = \beta S I - \mu I - \nu I\\
\dot{R} = \nu I - \mu R\\
(N = S + I + R)
\end{matrix}
\]
}

Now we modify the SIR example in order to include two therapies $T1$ and $T2$ that makes immune the susceptible population and increase the mutation rate from infected to recovered, respectively. $T1_{on}$ ($T2_{on}$) and $T1_{off}$ ($T2_{on}$) denote the therapy being administered or not, respectively. Therefore, $T1$ can be thought as a vaccination that makes immune the susceptible individuals, while $T2$ as a therapy for combating the course of the disease. The therapy-specific reactions introduced are listed in Table~\ref{tbl:sirmodel1}.
\begin{table}
\centering
\begin{small}
\begin{tabular}{rl}
$S + T1_{on}\rightarrow_\rho R + T1_{on}$&Susceptible becomes recovered with therapy 1\\
\noalign{\smallskip}
$I + T2_{on}\rightarrow_{k} R + T2_{on}$&Infected becomes recovered with therapy 2\\
\noalign{\smallskip}
$T1_{off} \rightarrow_{r1_{on}} T1_{on}$& Therapy 1 switched on\\
$T1_{on} \rightarrow_{r1_{off}} T1_{off}$& Therapy 1 switched off\\
$T2_{off} \rightarrow_{r2_{on}} T2_{on}$& Therapy 2 switched on\\
$T2_{on} \rightarrow_{r2_{off}} T2_{off}$& Therapy 2 switched off\\
\end{tabular}
\end{small}
\caption{Additional therapy-specific reactions in the SIR epidemic model with open population and medical treatments at Table~\ref{tbl:sirmodel}.}
\label{tbl:sirmodel1}
\end{table}
In this case, the associated D-CGF model $\mathcal{M}=(S, P, T, C)$ has been extended with the therapy processes and is defined as follows:
{\small
\[
S: \ \begin{array}{rl}
S = & \tau_{S_1}^b . (S \| S) \ + \ \tau_{S_2}^\mu . \mathbf{0}\ + \ ? i^\beta . I \ + \ ? j^\rho . R\\
I = & \tau_{I_1}^b . (I \| S) \ + \ \tau_{I_2}^\mu . \mathbf{0}\ + \ ? \tau_{I_3}^\nu . R \ + \ ! i^\beta . I\ + \ ? h^{k} . R\\
R = & \tau_{R_1}^b . (R \| S) \ + \ \tau_{R_2}^\mu . \mathbf{0}
\end{array}
\qquad
P:\ P_0
\]\[
T: \ \begin{array}{rl}
T1_{off} = & \tau_{1on}^{r1_{on}} . T1_{on}\\
T1_{on} = & ! j^\rho . T1_{on} \ + \  \tau_{1off}^{r1_{off}} . T1_{off}\\
T2_{off} = & \tau_{2on}^{r2_{on}} . T2_{on}\\
T2_{on} = & ! h^{\nu + k} . T2_{on} \ + \ \tau_{2off}^{r2_{off}} . T2_{off}\\
\end{array}
\qquad
C: \ T1_{off} \ \| \ T2_{off}
\]
}
\subsection{Hybrid semantics of D-CGF}\label{subsect:hybridsemantics}
In the following we will give the definition of switching therapy (ST), and show that STs constitute the discrete switches in the hybrid dynamical system semantics where the different species are the continuous variables. The definition of switching therapy is broadly inspired by the notion of Control Automata~\cite{bortolussi2009hybrid}, i.e.\ the control structure that can be identified in the hybrid automata-based semantics of a CGF model. We also provide a constructing procedure for determining the STs associated to a D-CGF model and therefore its semantics in terms of hybrid dynamical systems.


In order to interpret the set therapy terms $T$ as a set of discrete switches, we need to identify collections of terms in which exactly one term is active in every combination of therapies reachable from the initial one. We call such a collection a \textit{Switching Therapy (ST)}. The intuition is that a switching therapy models a discrete component whose terms represent its internal states, since exactly one of them must be active at each step. We give the definition of ST and of well-formed therapies for imposing additional restrictions on the construction of therapy terms in order to ensure a correct discrete interpretation of them.

\begin{defn}[Switching Therapy]\label{def:st}
Let $\mathcal{M} = (S,P,T,C)$ be a D-CGF model, and $T_i \subseteq T$ a set of therapy terms. $T_i$ is a \textit{Switching Therapy (ST)} if the following conditions hold:
\begin{enumerate}
\item Exactly one of the terms $U \in T_i$ is active in the initial combination of therapies: $\#(T_i, C) = \sum_{U \in T_i} \#(U, C) = 1$.
\item Each action must conserve the concentration of terms in $T_i$ and cannot involve more than one reagent in $T_i$: $\forall \pi \in \Pi. (\#(T_i, react(\pi))=\#(T_i, prod(\pi)) \leq 1)$.
\item For each action $\pi \in \Pi$ such that $U_1 \in react(\pi)$ and $U_2 \in prod(\pi)$, with $U_1, U_2 \in T_i$, $U_1 \neq U_2 \implies (react(\pi) = \lbrace U_1\rbrace \wedge prod(\pi) = \lbrace U_2\rbrace)$.
\end{enumerate}
\end{defn}
Conditions 1+2 ensure that exactly one term of the ST is active in every reachable configuration, since condition 1 requires that the initial concentration of $T_i$-terms must be equal to one and condition 2 tells that such concentration is conserved. Additionally, condition 2 implies that there are no actions able to modify the concentration of $T_i$-terms and thus that the internal state of the ST $T_i$ can be changed only by actions having a $T_i$-term in its reactants. Finally, condition 3 requires that every action causing the switch of a $T_i$-term from $U_1$ to $U_2$ must be an internal action of $U_1$, of the form $U_1 = \ldots + \tau.U_2 + \ldots$.
\begin{defn}[Well-formed therapies]\label{def:wellformed}
Let $\mathcal{M} = (S,P,T,C)$ be a D-CGF model and let $C=U_1 \| \ldots  \| U_n$, with $n \geq 1$. $T$ is a set of well-formed therapies if there exists a partition $\mathcal{T} = \lbrace T_1, \ldots , T_n \rbrace$ of $T$ such that $T_i \in \mathcal{T}$ is a switching therapy and $U_i \in T_i$, for all $i=1, \ldots, n$.
\end{defn}
We present a method for extracting the switching therapies and show how the semantics in terms hybrid dynamical systems can be constructed. Given a D-CGF model $\mathcal{M} = (S,P,T,C)$, the procedure exploits the stoichiometric matrix $M$, with which it is easy to check the conservation of switching terms and condition 3 of Definition~\ref{def:st}. Some necessary conditions has to be formulated on $M$ for ensuring the well-formedness of $T$.
\begin{defn}[Necessary conditions for well-formedness]\label{def:necessary}
Let $\mathcal{M} = (S,P,T,C)$ be a D-CGF model and $M$ be the associated stoichiometric matrix. Then, the following conditions are necessary for the well-formedness of $T$
\begin{enumerate}
\item The elements of $M$ restricted to $T$ take values in $\lbrace -1, 0, 1 \rbrace$: $\forall \pi \in \Pi. \ \forall U \in T. \ M_{|T}[U, \pi] \in \lbrace -1, 0, 1 \rbrace$.
\item Conservation of $T$-terms: $\forall \pi \in \Pi. \ \sum_{U \in T} M[U, \pi] \ = 0$;
\item Exclusive switch (1): $\forall \pi \in \Pi. \ \exists_{\leq 1} U \in T . \ M[U, \pi] = -1$. This condition and the conservation condition additionally imply that it exists exactly one $V \in T$ such that $M[V, \pi] = 1$.
\item Exclusive switch (2): $\forall \pi \in \Pi. \ (\exists U \in T . \ M[U, \pi] = -1 \implies (\forall X \in S . M[X, \pi]=0 \wedge \pi \text{ internal}))$. 
\end{enumerate}
\end{defn}

The stoichiometric matrix $M$ associated to the SIR example with therapies is given below.
{\footnotesize
\[
M = \bordermatrix{~ &\tau_{S_1} & \tau_{I_1} & \tau_{R_1} & \tau_{S_2} & \tau_{I_2} & \tau_{R_2} & \tau_{I_3} &\tau_{1on} & \tau_{1off} & \tau_{2on} & \tau_{2off} & i & j & h \cr
S& 1 & 1 & 1 & -1 & 0 & 0 &  0 & 0 & 0 & 0  &  0 & -1 & -1 & 0\cr
I& 0 & 0 & 0 & 0 & -1 & 0 &  -1 & 0 & 0 & 0  &  0 & 1 & 0 & -1\cr
R& 0 & 0 & 0 & 0 & 0 & -1 &  1 & 0 & 0 & 0  &  0 & 0 & 1 & 1\cr\cr
T1_{off}& 0 & 0 & 0 & 0 & 0 & 0 &  0 &-1 & 1 & 0  &  0 & 0 & 0 & 0\cr
T1_{on}& 0 & 0 & 0 & 0 & 0 & 0 &  0 & 1  & -1  & 0  &  0 & 0 & 0 & 0\cr
T2_{off}& 0 & 0 & 0 & 0 & 0 & 0 & 0 & 0  & 0  & -1 &  1 & 0 & 0 & 0\cr
T2_{on}& 0 & 0 & 0 & 0 & 0 & 0 &  0 & 0  & 0  & 1  & -1 & 0 & 0 & 0\cr}
\]
}

It is easy to check that the conditions listed above are met in $M$.
Such conditions on the stoichiometric matrix are not sufficient to ensure the well-formedness of the set of therapies $T$. Indeed, Def.~\ref{def:necessary} corresponds to stating that the whole set $T$ meets conditions 2+3 in the definition of switching therapy, but it is not possible from the only stoichiometric matrix to prove also condition 1 (exactly one term in $T$ is active in the initial configuration). Moreover recalling that $T$ is well-formed if it can be partitioned into a set of switching therapies, such necessary conditions are defined on the entire set of therapies $T$, thus considering only the trivial partition $\lbrace T \rbrace$.

The algorithm for extracting the STs consists in building a graph $\mathcal{G} = (T, E)$ called \textit{ST-graph}, where $T$ is the set of process terms and arcs connect couples $(U_1,U_2)$ of therapy terms that are involved in a switch, i.e.\ $M[U_1, \pi] = -1$ and $M_{|T}[U_2, \pi] = 1$ for some $\pi$.

\begin{defn}[ST-graph]
Let $\mathcal{M} = (S,P,T,C)$ be a D-CGF model and $M$ be the associated stoichiometric matrix. The ST-graph $\mathcal{G_M}=(T,E)$ associated to $\mathcal{M}$ is a directed graph whose vertices are the $T$-terms of $\mathcal{M}$ and $E=\lbrace (U_1,U_2) \in T \times T \ | \ \exists \pi \in \Pi . \ (M[U_1, \pi] = -1 \wedge M_{|T}[U_2, \pi] = 1)\rbrace$.
\end{defn}

Now it is possible to check the well-formedness of $T$ by considering the set of connected components of $\mathcal{G_M}$, $\mathcal{C(G_M)}$ (see Proposition~\ref{prop:wellform}). In particular, if the stoichiometric matrix $M$ meets the necessary conditions in Def.~\ref{def:necessary} and the set of connected components $\mathcal{C(G_M)}=\lbrace G_1, \ldots, G_n\rbrace$ is such that for each $G_i$ there is exactly one term active in the initial configuration, then $T$ is well-formed and the nodes of each $G_i$ form a switching therapy.
Note that the sets of nodes of the connected components in an ST-graph form, by definition, a partition of $T$. This is necessary to guarantee that, according to Def.~\ref{def:wellformed}, a set well-formed therapies can be partitioned into a set of switching therapies. In addition, the therapy terms in each $G_i$ meet the Condition 2 of Def.~\ref{def:st} about the conservation of terms in a switching therapy, because the connected components $G_i$ are (of course) mutually disconnected, i.e.\ there cannot exist any arc in the ST-graph connecting them that is, no terms can flow between them.

\begin{prop}\label{prop:wellform}
Let $\mathcal{M} = (S,P,T,C)$ be a D-CGF model, $M$ be the associated stoichiometric matrix, $\mathcal{C(G_M)} = \lbrace G_1 = (T_1,E_1), \ldots , G_n  = (T_n,E_n) \rbrace$ be the set of connected components of the associated ST-graph $\mathcal{G_M}$ and $\mathcal{T}=\lbrace T_1, \ldots, T_n \rbrace$. $T$ is well-formed and each $T_i \in \mathcal{T}$ is a switching therapy if the following conditions hold:
\begin{enumerate}
\item $M$ meets the necessary requirements at Def.~\ref{def:necessary};
\item for each $T_i \in \mathcal{T}$, exactly one of the terms in $T_i$ is active in the initial combination of therapies: $\#(T_i, C) = 1$; and
\item for each $T_i \in \mathcal{T}$ and for each action $\pi \in \Pi$, the number of $T_i$-terms in $react(\pi)$ is at most one: $\#(T_i, react(\pi)) \leq 1$.
\end{enumerate}
\end{prop}

\begin{figure}
\centering
\subfloat[]{\includegraphics[height=3cm]{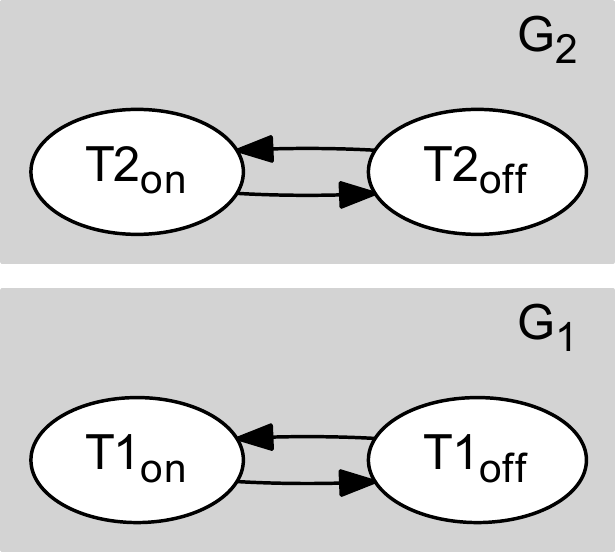}}\hspace{1cm}
\subfloat[]{\includegraphics[height=3cm]{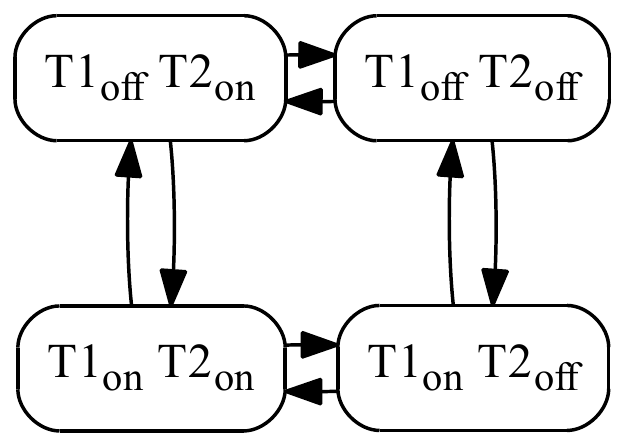}}
\caption{ST-graph for the SIR example (a) and its corresponding mode graph (b). The connected components of the ST-graph are such that there is exactly one term in each $G_i$ active in $C$ and for each $G_i$, no action $\pi$ involves more than one $G_i$-term in $react(\pi)$. This ensures that $T$ is well formed and the connected nodes are switching therapies. In (b) four distinct discrete modes are generated from the Cartesian product of $G_1$ and $G_2$.}
\label{fig:stGraph}
\end{figure}

The ST-graph for the SIR example is given in Fig.~\ref{fig:stGraph} (a). It is straightforward to check that its connected components are such that exactly one term in each $G_i$ is active in the initial combination of therapies $C = T1_{off} \ | \ T2_{off}$; and that for each $G_i$ there is no action $\pi$ whose reactants involve more than one term in $G_i$. Therefore the set of therapy definitions $T$ of the SIR example is well-formed and $\mathcal{T} = \lbrace \lbrace T1_{off}, T1_{on}\rbrace, \lbrace T2_{off}, T2_{on}\rbrace \rbrace$ is the set of switching terms.

The combination of the different switching therapies allows us to extract the possible discrete \textit{modes} in the hybrid semantics. We define a \textit{mode graph} structure as the Cartesian product of the switching terms in a ST-graph. Figure~\ref{fig:stGraph} (b) shows the mode graph for the SIR example.
\begin{defn}[Mode graph]
Let $\mathcal{M} = (S,P,T,C)$ be a D-CGF model and $\mathcal{C(G_M)} = \lbrace G_1, \ldots , G_n \rbrace$ be the set of connected components of the associated ST-graph. The mode graph $MG_\mathcal{M}$ is defined by the Cartesian product $G_1 \times \ldots \times G_n$.
\end{defn}
Note that transitions between modes could have associated exponentially distributed delays determined by the stochastic rates of the algebraic specification. In our context, we omit them assuming that such transitions are instantaneous, also because discrete modes will represent the control inputs of the hybrid dynamical system semantics.

As pointed out also in~\cite{bortolussi2009hybrid}, Petri Nets could have been similarly applied in order to study the conservation properties in a switching therapy. In particular, a set of switching therapies could be seen as a set of \textit{strictly conservative} (i.e.\ constant number of tokens) and \textit{1-bounded} (i.e.\ the maximum number of token is 1) Petri Nets.
\subsubsection{Hybrid dynamical system of D-CGF models}
\begin{figure}
\centering
\includegraphics[width=0.5\textwidth]{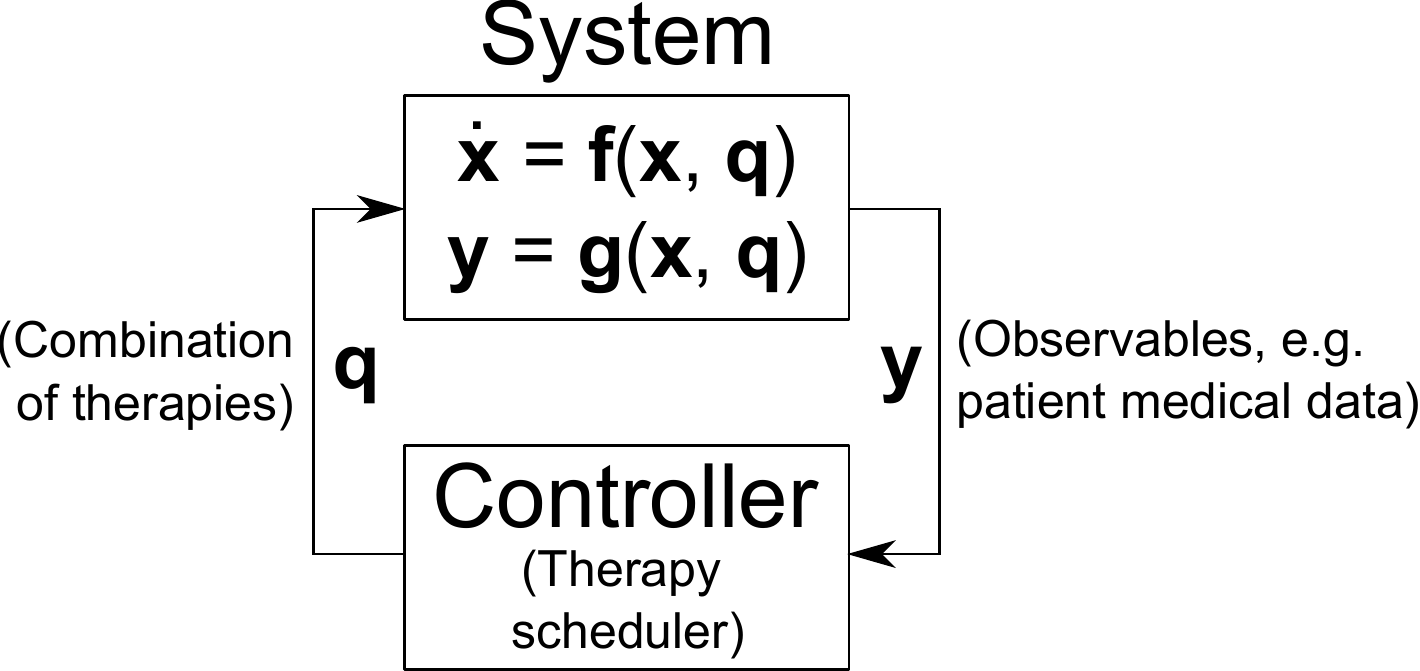}
\caption{Control loop in a controlled switched system. The plant defines the piecewise smooth dynamics of the system $\mathbf{\dot{x}}$ and the observable output $\mathbf{y}$. The output could represent patient's medical data obtained after a visit. Depending on $\mathbf{y}$, the controller acts as therapy scheduler and determines the operation mode $\mathbf{q}$ (e.g.\ a particular combination of therapies) of the plant.}
\label{fig:css_control_loop}
\end{figure}

Here we describe the semantics of a D-CGF model in terms of a particular class of hybrid dynamical systems, called \textit{controlled switched systems (CSS)}~\cite{lunze2009handbook}. While in the general formulation of hybrid dynamical system the external control input and the discrete operation mode are distinct, in a CSS the external controller produces a switching signal (i.e.\ the discrete mode) that is given in input to the plant (i.e.\ the controlled system). The basic form of a controlled switched system is the following:
\begin{align*}
\mathbf{\dot{x}} = & \mathbf{f}(\mathbf{x}, \mathbf{q})\\
\mathbf{y} = & \mathbf{g}(\mathbf{x}, \mathbf{q}),
\end{align*}
where $\mathbf{x} \in \mathbb{R}^n$ is the continuous state, $\mathbf{f}$ is a vector-valued smooth function, $\mathbf{g}$ is the output function and $\mathbf{q}$ is the discrete operation mode. The above equations are often written with the form $\mathbf{\dot{x}} = \mathbf{f_q}(\mathbf{x}) \quad \mathbf{y} = \mathbf{g_q}(\mathbf{x})$, stressing the different dynamics and outputs under different operation modes. Note that a controlled switched system can also be seen as a hybrid dynamical system with controlled discrete inputs. Figure~\ref{fig:css_control_loop} shows the control loop in a CSS.
We show how to derive a CSS from a D-CGF model, by following the method elaborated by Cardelli for the CGF algebra as regards the continuous part of the semantics, but considering also the switching therapies and the mode graph as regards  the discrete part. Recalling that each node of a mode graph $MG_\mathcal{M}=(\lbrace q_1=\lbrace U_{11}, \ldots, U_{1n}\rbrace, \ldots , q_k=\lbrace U_{k1}, \ldots, U_{km}\rbrace\rbrace, E)$ models a particular combination of therapies, a mode $q_i$ being active means that therapies $U\in q_i$ are active and that those in $T \setminus q_i$ are not. Therefore a modified version of the rate vector will be used: $$\phi_{q_i} = \phi[\faktor{\mathbf{1}}{U},\faktor{\mathbf{0}}{V}], \ U \in q_i, V \in T \setminus q_i,$$ which is obtained by zeroing all the elements of $\phi$ that contain a therapy term not belonging to $q_i$, and omitting (substitution with 1) those in $q_1$ . The resulting controlled switched system would be:
\[
\mathbf{\dot{x}} = \mathbf{f}_{q_i}(\mathbf{x}) = M_{|S} \cdot \phi_{q_i},
\]
where $q_i \in MG_\mathcal{M}$ is a node of the mode graph, and $M_{|S}$ is the stoichiometric matrix restricted to the set of species $S$. The operation modes $\mathbf{q}$, the modified rate vector $\phi_{\mathbf{q}}$ and the resulting CSS $\mathbf{\dot{x}}$ of the SIR example are shown below.
{\footnotesize
\[
\phi_{\mathbf{q}} = \bordermatrix{~ &\tau_{S_1} & \tau_{I_1} & \tau_{R_1} & \tau_{S_2} & \tau_{I_2} & \tau_{R_2} & \tau_{I_3} & \tau_{1on} & \tau_{1off} & \tau_{2on} & \tau_{2off} & i & j & h \cr
q_1& bS     & bI     & bR     & \mu S  & \mu I  & \mu R  &  \nu I  & r1_{on} & 0 & r2_{on}  &  0 & \beta SI & 0 & 0 \cr
q_2& \vdots & \vdots & \vdots & \vdots & \vdots & \vdots &  \vdots  & 0 & r1_{off} & r2_{on} &  0 & \vdots & \rho S &  0\cr
q_3& \vdots & \vdots & \vdots & \vdots & \vdots & \vdots &  \vdots  & r1_{on} & 0 & 0 & r2_{off} & \vdots & 0 &  k I\cr
q_4& bS     & bI     & bR     & \mu S  & \mu I  & \mu R  &  \nu I  & 0 & r1_{off} & r2_{off} & 0 & \beta SI & \rho S & k I\cr}\]\[
\mathbf{q}=\begin{bmatrix}
q_1 = (T1_{off}, T2_{off})\\
q_2 = (T1_{on}, T2_{off})\\
q_3 = (T1_{off}, T2_{on})\\
q_4 = (T1_{on}, T2_{on})
\end{bmatrix}
\qquad
\mathbf{\dot{x}}(q_1)= \left\{ 
  \begin{array}{l}
	\dot{S} = bN - \beta S I - \mu S\\
	\dot{I} = \beta S I - \mu I - \nu I\\
	\dot{R} = \nu I - \mu R\\
  \end{array} \right.
  \qquad
\mathbf{\dot{x}}(q_2)= \left\{ 
  \begin{array}{l}
	\dot{S} = bN - \beta S I - \mu S - \rho S\\
	\dot{I} = \beta S I - \mu I - \nu I\\
	\dot{R} = \nu I - \mu R + \rho S\\
  \end{array} \right.
\]\[
\mathbf{\dot{x}}(q_3)= \left\{ 
  \begin{array}{l}
	\dot{S} = bN - \beta S I - \mu S \\
	\dot{I} = \beta S I - \mu I - (\nu + k) I\\
	\dot{R} = (\nu + k) I - \mu R\\
  \end{array} \right.
    \qquad
\mathbf{\dot{x}}(q_4)= \left\{ 
  \begin{array}{l}
	\dot{S} = bN - \beta S I - \mu S - \rho S\\
	\dot{I} = \beta S I - \mu I - (\nu + k) I\\
	\dot{R} = (\nu + k) I - \mu R + \rho S\\
  \end{array} \right.
\]
}\noindent Alternatively, we could consider the equivalent formulation which exploits the sets of switching terms $\lbrace T1_{off}, T1_{on}\rbrace$ and $\lbrace T2_{off}, T2_{on}\rbrace$ as discrete control variables in the following way.
{\footnotesize
\[
T_1(q_i)= \left\{ 
  \begin{array}{ll}
	1 & \quad \text{if } T1_{on} \in q_i\\
	0 & \quad \text{if } T1_{off} \in q_i\\
  \end{array} \right.
  \qquad
  T_2(q_i)= \left\{ 
  \begin{array}{ll}
	1 & \quad \text{if } T2_{on} \in q_i\\
	0 & \quad \text{if } T2_{off} \in q_i\\
  \end{array} \right.
  \qquad
  \mathbf{\dot{x}}(q_i)= \left\{ 
  \begin{array}{l}
	\dot{S} = bN - \beta S I - \mu S - T_1(q_i)\rho S\\
	\dot{I} = \beta S I - \mu I - (\nu + T_2(q_i)k) I\\
	\dot{R} = (\nu + T_2(q_i)k) I - \mu R + T_1(q_i)\rho S\\
  \end{array}\right.
\]
}\noindent Note that this form is applicable because in the SIR example each switching term contains exactly two terms and can consequently have assigned a value in $\lbrace 0,1 \rbrace$.
\section{Hybrid semantics and optimal control of infectious processes}\label{sect:sim}
In this section we define the optimal control law for therapy scheduling in the SIR model by means of Model Predictive Control (MPC)~\cite{bemporad1999control}, and we report the optimally controlled solutions with parameters that are typical of the measles~\cite{stone2000theoretical}. A similar problem has been addressed in~\cite{suzuki2010piecewise} for the scheduling of hormone therapy in a model for prostate cancer.
%
%

In order to cope with the nonlinear dynamics of the SIR model, the optimal scheduling of multiple therapies is computed by solving an on-line MPC problem in the Multi-Parametric Toolbox (MPT)~\cite{mpt}. The optimal control law can be formulated by means of the following constrained finite-time optimal control (CFTOC) problem, where the original continuous-time dynamics is simply discretized with the Euler method:
{\footnotesize
\[\begin{array}{ll}
&\underset{\mathbf{u}}{min} \sum_{k=0}^{T-1} \Vert R\mathbf{u}(k) \Vert_{1} + \Vert Q\mathbf{x}(k) \Vert_{1}\\
subj. \ to \quad & \mathbf{x}(k) \in [0,1]^3\\
& \mathbf{u}(k) \in \lbrace 0,1\rbrace^2\\
& \mathbf{x}(T) \in \mathcal{X}_{set}\\
&\mathbf{x}(k+\Delta t)= A\mathbf{x}(k) + B\mathbf{u}(k)\\
&\forall k = 0,\ldots , T
\end{array}
\quad
\begin{array}{ll}
A = &\begin{bmatrix}
1 + (b -\beta I(k)-\mu)\Delta t & b\Delta t & b\Delta t\\
0 & 1 + (\beta S(k)-\mu - \nu)\Delta t & 0\\
0 & \nu\Delta t & 1 -\mu\Delta t
\end{bmatrix}\\
B = &\begin{bmatrix}
-\rho S(k) \Delta t & 0\\
0 & -k I(k) \Delta t\\
\rho S(k) \Delta t & k I(k) \Delta t\end{bmatrix} \quad 
Q = \begin{bmatrix} 1 & 0 & 0\\ 0 & 10 & 0\\ 0 &0&0.5\end{bmatrix}
\end{array}
\]
}\noindent where $\Delta t = 1/365$ is the discrete time step corresponding to one day; $\mathbf{u}(k)= [T1(k) \ T2(k)]^T \in \lbrace 0,1\rbrace^2$ is the discrete input (the combination of the two therapies) and $\mathbf{x}(k)= [S(k) \ I(k) \ R(k)]^T \in [0,1]^3$ is the systems' continuous state vector at time $k$; $T=3$ is the prediction horizon (corresponding to three days); $\mathcal{X}_{set}$ is a polytope identifying the set of terminal states; $\Vert \mathbf{x} \Vert_{1} = |S| + |I| + |R|$ and $\Vert \mathbf{u} \Vert_{1} = |T1| + |T2|$; $R$ and $Q$ are the weights on the controlled inputs and on the system state, resp. The terminal set $\mathcal{X}_{set}$ represents an invariant on the final states and can be described by the convex hull of its vertices: $\mathcal{X}_{set} = Conv(\lbrace [1 \ 0 \ 0]^T, [0 \ 0 \ 1]^T\rbrace)$. In this way, the terminal set contains all the states $[S \ I \ R]^T$ such that $I=0$ (no infected individuals), $S,R \geq 0$ and $S+R = 1$ (we do not exceed the population limit). Matrix $Q$ has been set so that the state variable Infected has associated a weight equals to $10$ and thus is slightly penalized over variables Susceptible (weight $1$) and Recovered (weight $0.5$). Finally, we assume for simplicity that the internal state can be observed and we set the output vector $\mathbf{y}$ to the state vector $\mathbf{x}$.

Matrix $R$ allows us to implement different therapy scheduling strategies. The higher $R$, the higher penalty is given to drug dosage. In particular if $R$ is assigned a large value, then the strategy is to avoid drug dosage as much as possible. This can be the case of a therapy with severe side effects or of a patient in an early stage of the disease that can be treated even with a small dosage. On the contrary, a low value to $R$ leads to a strategy where drug dosage is much more prominent (e.g.\ little side effects or mature stage of disease). Additionally we can assign different weights to different drugs according to their therapeutic impact on the patient, thus possibly enabling a wider spectrum of control strategies. We computed the optimally controlled solutions considering three scenarios obtained by varying the weights on the controlled therapies:
\begin{enumerate}
\item Low penalties to $T1$ and $T2$: $R=\begin{bmatrix} 0.1 & 0 \\ 0 & 0.1\end{bmatrix}$
\item High penalty to $T1$: $R=\begin{bmatrix} 100 & 0 \\ 0 & 0.1\end{bmatrix}$
\item High penalty to $T2$: $R=\begin{bmatrix} 0.1 & 0 \\ 0 & 100\end{bmatrix}$
\end{enumerate}

\begin{figure}
\centering
\subfloat[Scenarios 1,2]{\includegraphics[width=0.3\textwidth, height=0.5\textwidth]{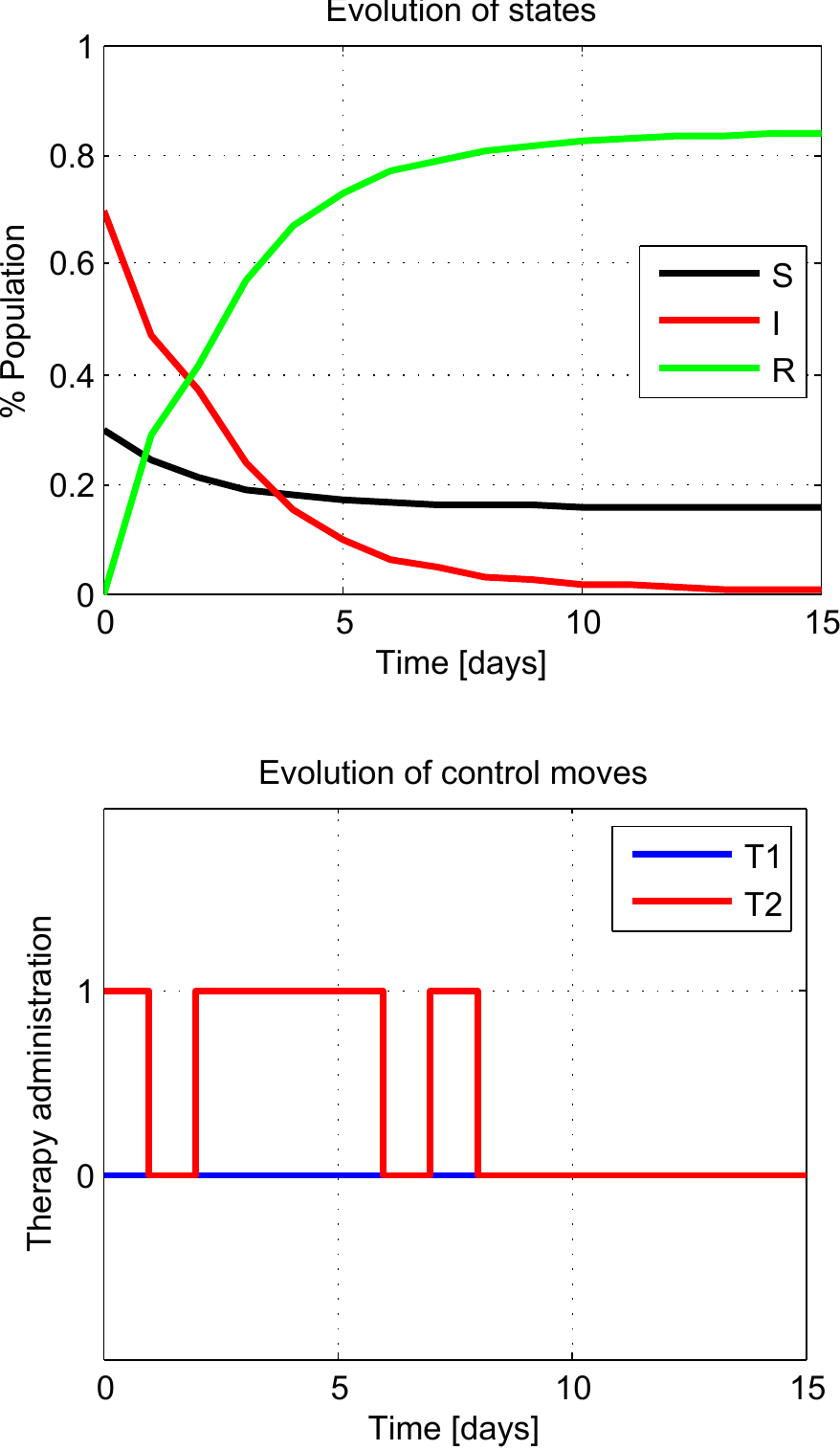}}\hspace*{1cm}
\subfloat[Scenario 3]{\includegraphics[width=0.3\textwidth, height=0.5\textwidth]{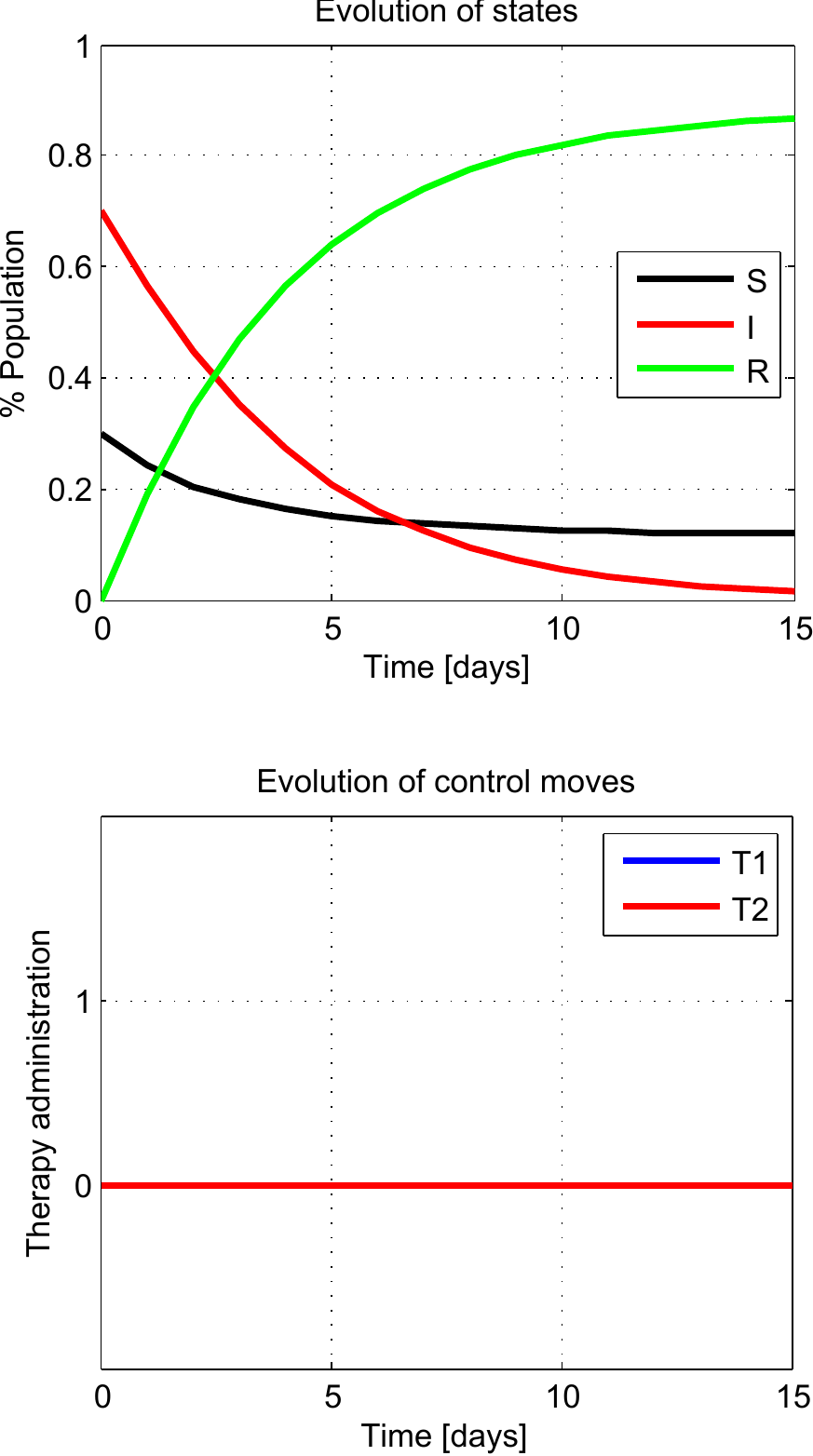}}
\caption{Optimally controlled solutions for the SIR model with therapies as controlled discrete inputs simulated for 15 days. Parameters are: $S(0)=0.3, \ I(0)=0.7, \ R(0)=0,\ b = \mu = 0.02, \ \beta=1800, \ \nu = 100, \ \rho=0.5, \ k=50$. Results have been computed by using a global optimal nonlinear solver.}
\label{fig:SIR_MPT}
\end{figure}

Results reported in Figure~\ref{fig:SIR_MPT} tell that solutions under scenario 1 and scenario 2 are the same (Fig.~\ref{fig:SIR_MPT} (a)), suggesting that therapy $T1$ does not need to be administered in the optimal strategy of treatment in order to fulfil the terminal constraints. Indeed Fig.~\ref{fig:SIR_MPT} (b) shows that when $T2$ has a high weight (scenario 3), no control moves are performed, so indicating that $T1$ is ineffective with respect to $T2$.

\subsection{Osteomyelitis, a truly multiscale infection}
Osteomyelitis is a a bone pathology caused by bacteria infection (mostly \textit{Staphylococcus aureus}) that alters the bone remodelling process and that rapidly leads to severe bone loss, necrosis of the affected portion, and it may even spread to other parts of the body. Therefore it is one of the most appropriate examples of multiscale infection, since it affects multiple biological scales. Indeed infection by \textit{S. aureus} starts at the intracellular level, but it subsequently involves also the tissue and the organ level. \textit{Bone remodelling (BR)} is the cellular-level process by which the bone is continuously renewed as the result of an alternation of bone resorption, conducted by cells called {\it osteoclasts}, and bone formation, conducted by {\it osteoblasts}. They form together the so-called \textit{Basic Multicellular Unit (BMU)}, i.e.\ an ensemble of osteoclasts and osteoblasts that dissolve an area of the bone surface and then fill it. The BMU could be considered as an emerging scale intermediate between the cell and tissue levels. Osteomyelitis induces a severe inflammatory response followed by progressive bone destruction and loss of the vasculature and with a persistent chronic infection; this is further complicated by the rapid emergence of resistant strains of \textit{S. aureus}. It has been shown that the infection prevents proliferation, induces apoptosis and inhibits mineralisation of cultured osteoblasts. Although effective treatment of this disease is very difficult, one of most used drug is the fusidic acid that acts as a bacteriostatic agent, and is usually combined with other antibiotics. Based on a recent work of the authors about the modelling of osteomyelitis and the comparison of different treatments~\cite{lio2012osteomyelitis}, we can define a hybrid version of that model and a treatment strategy combining antibiotic and anti-inflammatory therapies. The model describes how the bone remodelling dynamics of osteoblasts' ($Ob$) and osteoclasts' ($Oc$) population is affected by the \textit{S.aureus} ($B$).
{\small
\[
\mathbf{x}=\begin{bmatrix}
O_{c}\\O_{b}\\B
\end{bmatrix}\quad
\mathbf{u}=\begin{bmatrix}
T1\\T2
\end{bmatrix} \quad
\begin{matrix}
\dot{O_{c}}=\alpha_{1}O_{c}^{g_{11}(1+f_{11}\frac{B}{s})}O_{b}^{g_{21}(1 +T2\cdot k_i- f_{21}\frac{B}{s})}-\beta_{1}O_{c},\\
\dot{O_{b}}=\alpha_{2}O_{c}^{g_{12}/(1+f_{12}\frac{B}{s})}O_{b}^{g_{22}-f_{22}\frac{B}{s}}- \beta_{2} O_{b},\\
\dot{B}=(1-T1)\gamma_{B}B\cdot log(\frac{s}{B})
\quad
y = -k_1\cdot Oc + k_2 \cdot Ob,
\end{matrix}
\]
}\noindent where $\alpha_{i}$ and $\beta_{i}$ are growth and death rates; $g_{ij}$ describes the effectiveness of autocrine and paracrine regulation, i.e.\ the chemical interactions between osteoblasts and osteoclasts; $f_{ij}$ models the impact of the infection on the autocrine and paracrine regulation; $T1$ is the discrete input modelling the antibiotic dosage; $T2$ models the dosage of a anti-inflammatory therapy; $k_i$ is the 
effectiveness of therapy $T2$; the bacterial population follows a Gompertz curve with carrying capacity $s$ and growth rate $\gamma_{B}$; and the output function $y$ represents the bone density calculated as a function of $Oc$, $Ob$ and their resorption ($k_1$) and formation ($k_2$) rates. Bone Mineral Density (BMD) measurement are often taken in medical practice, thus $y$ is the quantity observed by the controller (a doctor who administers the treatment) after each visit.

The optimal scheduling of bacteriostatic and anti-inflammatory therapies could be formulated with a control law similar to the SIR example. Since the combination of antibiotic and anti-inflammatory drugs has never been applied in medical practice, we could set the weight matrix of input therapies $R$ so that $R_{12}$ and $R_{21}$ are much larger that $R_{11}$ and $R_{22}$. In this way, we give a high penalty to the combination of therapies $T1$ and $T2$ w.r.t. $T1$ and $T2$ taken individually.

\section{Discussion and conclusion}\label{sect:conclusions}
The field of predictive models in biomedicine is challenged by the need of a better comprehension of the phenomenon of scales in the biological organisation, particularly their role in the transition between health and disease conditions. The multiscale modelling of molecules-cell-tissue-organ-body interactions is a key step in the process of identifying the most important parameters acting in a disease state and their calibration, linking basic research and clinical practice (therapies). Here we discuss how a description based on hybrid dynamical systems is beneficial to the formulation of a multiscale modelling in biomedical processes. A second instance of the utility of hybrid dynamics approach stems from the presence of multiple controls in biological systems. These controls could be framed as occurring naturally (the immune system response) or induced after the administration of a therapy. It is very often the case of multiple treatments, i.e. switching between therapies or combination of therapies which is the case considered in this study. In particular, we show the suitability of the D-CGF high-level process algebra to accomplish the task: here, its semantics is given in terms of a hybrid dynamical system where the large number of cell populations provides the basis for a continuous modelling approach and the dosage of multiple therapies is implemented as discrete switches.

We believe that this approach opens several interesting directions in basic and clinical research.
First, it is general and could be used for different tissues and organs or multiorgan diseases. Second, it is possible to extend the mathematical formulation to all the scales of biological organisation involved in the infection and recovery conditions. Third, the therapy could be complex and built on several nested and hierarchical protocols.
One final comment: given the richness of examples provided by biological processes and medical therapies, there is basis for deriving interesting theories. We remind the \textit{bon mot} of Stan Ulam: Ask not what mathematics can do for biology. Ask what biology can do for Mathematics.

\bibliographystyle{eptcs}
\bibliography{hsb}
\end{document}